\documentclass[a4,11pt,twoside]{article}
\usepackage{graphicx}
\begin{document}
\title{Study of Backgrounds at JLC IR
       \thanks{
%This work is supported in part by Grant-in-Aid from the Ministry of Education,
%Culture, Sports, Science and Technology of Japan
Presented at the International Workshop on Linear Colliders, August 26 - 30, 2002 in 
Jeju, Korea.
}} 
\author{H.~Aihara~\thanks{e-mail address: aihara@phys.s.u-tokyo.ac.jp},   
	M.~Iwasaki and K.~Tanabe
\\
\\
        {\it Department of Physics, University of Tokyo, Tokyo, Japan}
}
\date{}
\maketitle
\begin{abstract}
A full simulation program based on GEANT4 has been developed to study
beam-induced backgrounds in the JLC beam delivery system. 
We report some results obtained using this program.
\end{abstract}

\section{JLC Beam Delivery System}

\begin{figure}[htb]
\begin{center}
\resizebox{0.8\textwidth}{!}{\includegraphics{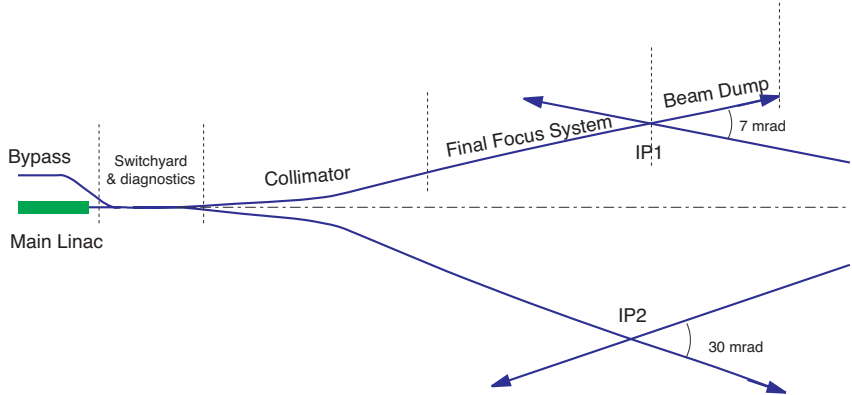}}
 \caption{\label{fig:BDSlayout}Schematic plan of the beam delivery  
section.}
\end{center}
\end{figure}

The beam delivery system (BDS) of the JLC  is $\sim1.4$~km long and   consists of four parts: switch-yard,
collimator, final focus system (FFS), and beam dump as shown in Fig.~\ref{fig:BDSlayout}.
\begin{figure}[htb]
\begin{center}
\resizebox{\textwidth}{!}{\includegraphics{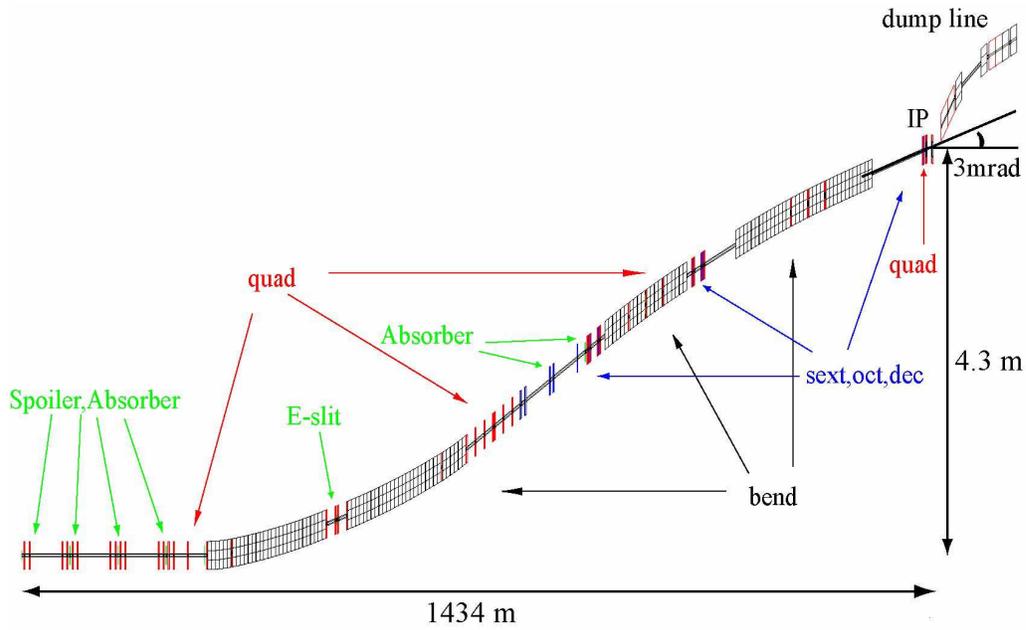}}
  \caption{\label{fig:BDS_JLC} Geometry of JLC beam delivery system. Aspect ratio between vertical and
horizontal scales is highly distorted.}
\end{center}
\end{figure}
\begin{figure}[h]
\begin{center}
\resizebox{0.8\textwidth}{!}{\includegraphics{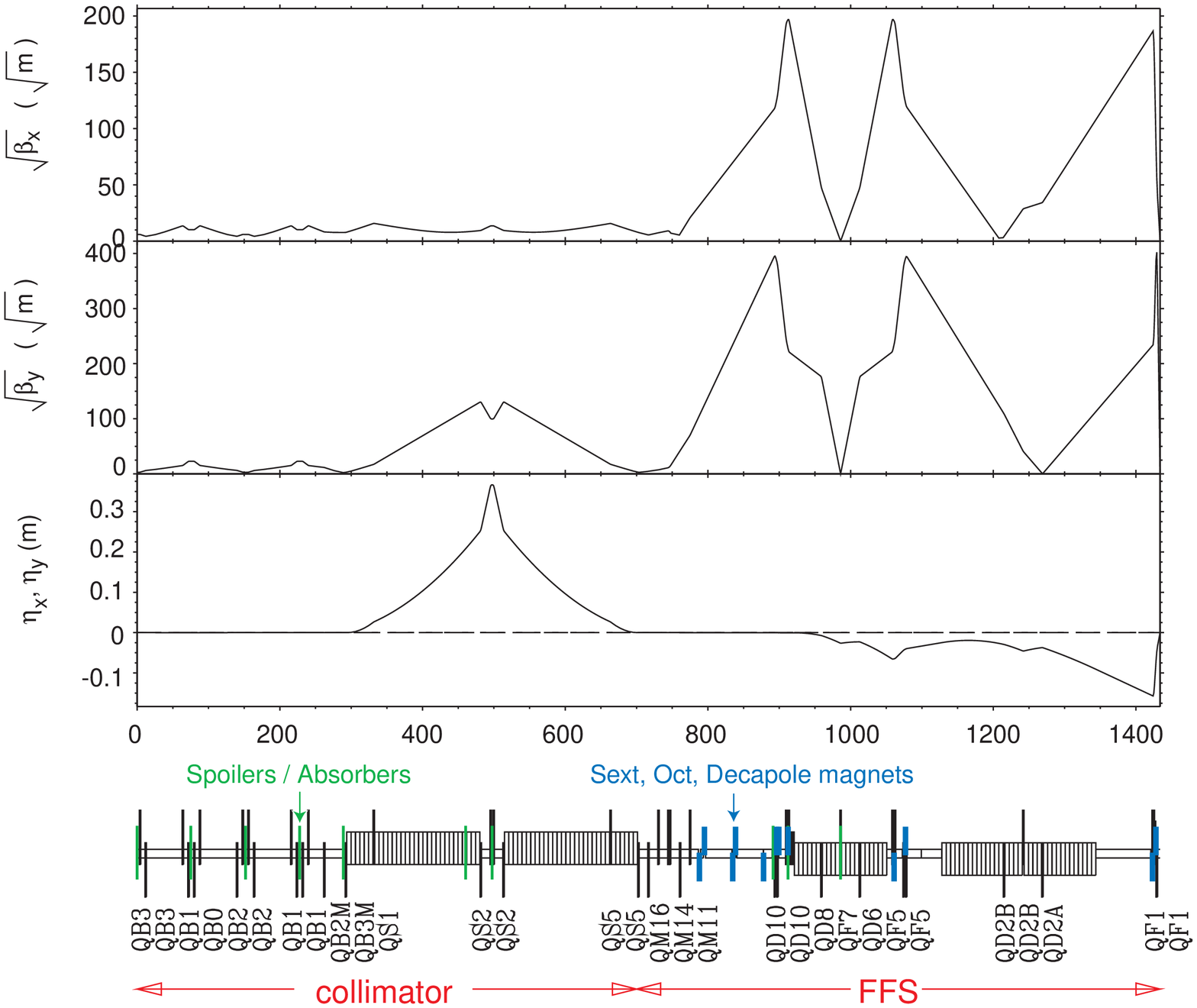}}
  \caption{\label{fig:BDSoptics}The optics functions of the beam delivery section.}
\end{center}
\end {figure}
The BDS serves multiple purposes: (i) to switch beams from the main linac or 
the bypass line to the first or second interaction point (IP),
(ii) to create a finite crossing angle  at the IP, 
(iii) to collimate the beams to  eliminate possible backgrounds at the detector,
(iv) to focus the beams at the IP,
(v) to protect the machine from damages due to beam aborts, 
and (vi) to dump the spent beams after collision safely.
The design of the JLC BDS is based on that of NLC BDS,
because of the close similarity of the overall machine design and
parameters. 
A notable difference, however,
exists between  JLC and NLC in a layout of the BDS.  
It arises
from the beam crossing angle at the first IP; 
JLC has  a small
angle of 7~mrad while NLC has 20~mrad.  
Figure~\ref{fig:BDS_JLC} shows
the geomtry of a JLC BDS.  The aspect ratio in this
diagram is highly distorted to illustrate the bending of the beam
lines and the beam crossing angle at the IP. 
Figure~\ref{fig:BDSoptics}
shows the optics functions in the collimator and final focus sections.
In the following we assume the electron beam energy of 250~GeV.

\section{BDS Simulation}
\begin{figure}[htb]
\begin{center}
\resizebox{0.8\textwidth}{!}{\includegraphics{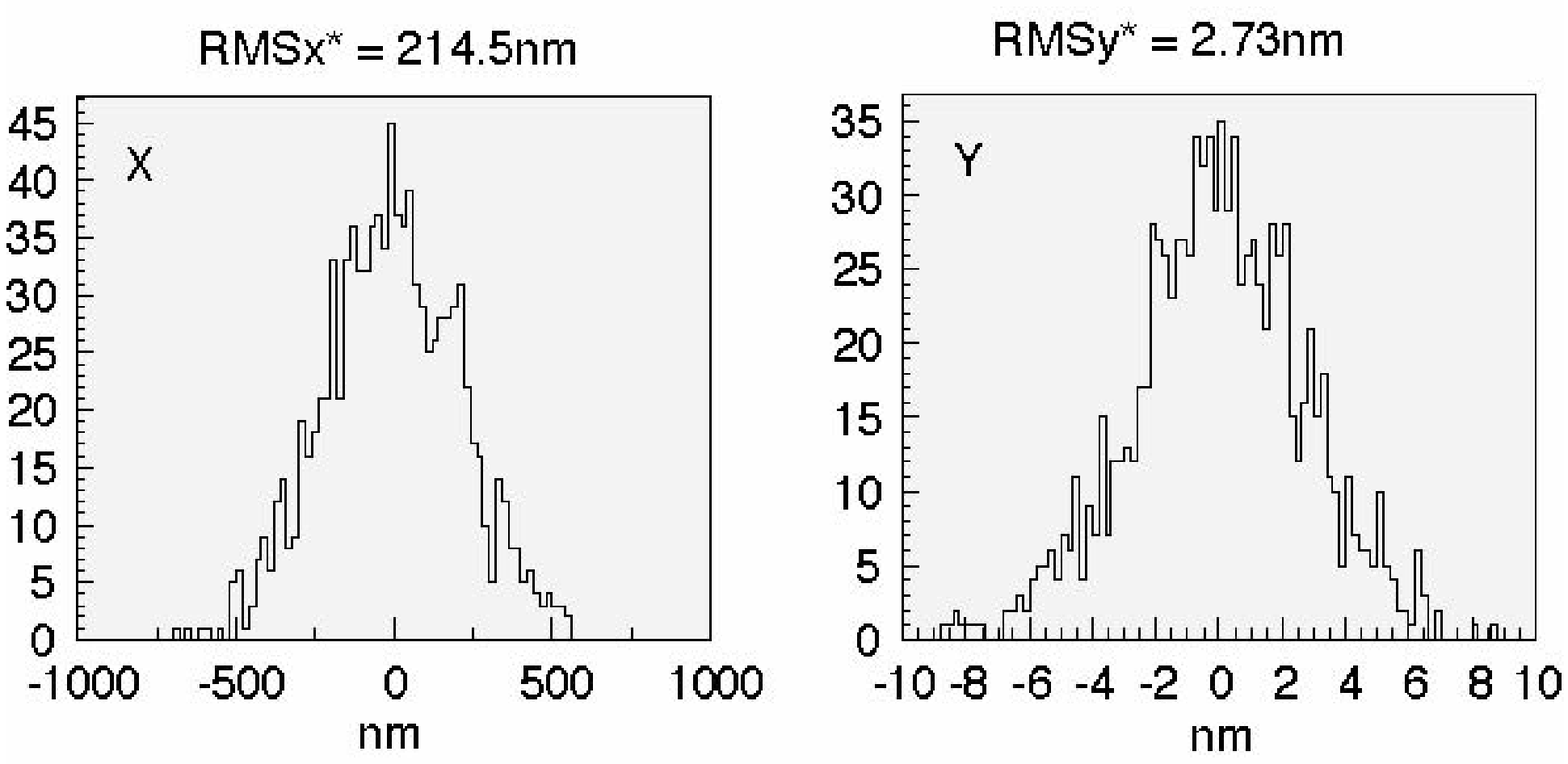}}
  \caption{\label{fig:size} The beam size at the IP, obtained by the LCBDS.}
\end{center}
\end{figure}

A full simulation program, LCBDS~\cite{LCBDS},   has been
developed to simulate the entire BDS.
%to study possible backgrounds and
%to investigate the countermeasures.  
The LCBDS 
uses GEANT4~\cite{GEANT4} to model the individual beamline
elements including drifts, all (dipole, quadrupole, sextupole,
octapole, and  decapole) magnets, collimator elements (spoiler, absorber,
and energy slit), and a beampipe.  Physics processes currently included in
the code are multiple scattering at the beamline elements,
electromagnetic shower and photon conversion to muon pairs.  In order
to gain reasonable statistics, we increase the muon production cross
section by a factor of 100 to generate events and, then, scale down
the results by the same factor.
In order to verify the tracking capability of the simulation, the transportation of the core
beam has been simulated from the entrance of the BDS to the IP
and the results are compared with the
beam profile shown in Fig.~\ref{fig:BDSoptics} obtained by using SAD program~\cite{SAD}.
The core beam is assumed to have a size of $\sigma_x=
14.8~\mu{\rm m} $ and $\sigma_y = 658~{\rm nm}$, and an angular
dispersion of $\sigma_{x^\prime}= 0.413~\mu{\rm rad} $ and
$\sigma_{y^\prime}= 0.093~\mu{\rm rad} $ at the entrance of the BDS.
Figure~\ref{fig:size} shows the beam size at the IP obtained by the LCBDS.
The results  (214.5 nm and 2.73 nm in $x$ and $y$ directions,
respectively, in rms) well reproduce the design values (211~nm and
2.7~nm).

\section{Detector Backgrounds}
The sources of the background can be categorized as follows:
\begin{enumerate}\itemsep 0mm
\item Upstream background: background produced upstream of the IP.  
This includes synchrotron radiation generated
at the final focus quadrupole and bending magnets, and muons produced
in electromagnetic showers arising from lost halo electrons/positrons.
Because the collimators are designed to scrape off halo particles,
muons are produced predominantly at the collimators.
\item IP background: background generated at the IP. 
Produced via $e^+e^-$
collisions at the IP are disrupted primary beams, beamstrahlung photons,
$e^+e^-$ pairs from beam-beam interactions, radiative Bhabhas and
hadrons from two-photon interactions.
\item Downstream background: background produced downstream of the IP.
This concerns secondary neutrons produced at the extraction line and
beam dump.  A large number of low-energy neutrons can drift back and
back-shine the detector.
\end{enumerate}
We report  on-going
efforts to identify background sources based on the LCBDS simulation.

\subsection{Synchrotron Radiation Background}
\begin{figure}[htb]
\begin{center}
\resizebox{0.6\textwidth}{!}{\includegraphics{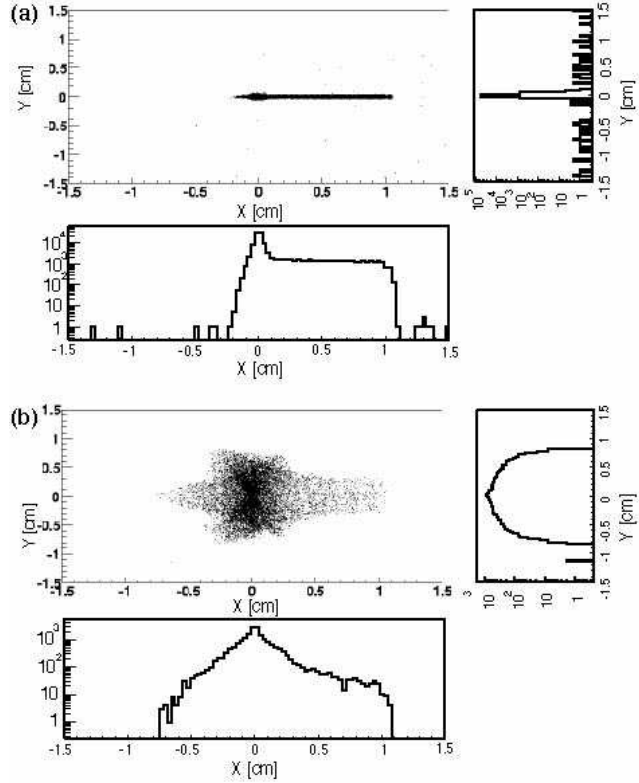}}
  \caption{\label{fig:sor_space} The spatial distributions of the synchrotron 
radiation photons arising from  (a) a nominal beam and (b) a beam halo .}
\end{center}
\end{figure}

\begin{figure}[htb]
\begin{center}
\resizebox{0.5\textwidth}{!}{\includegraphics{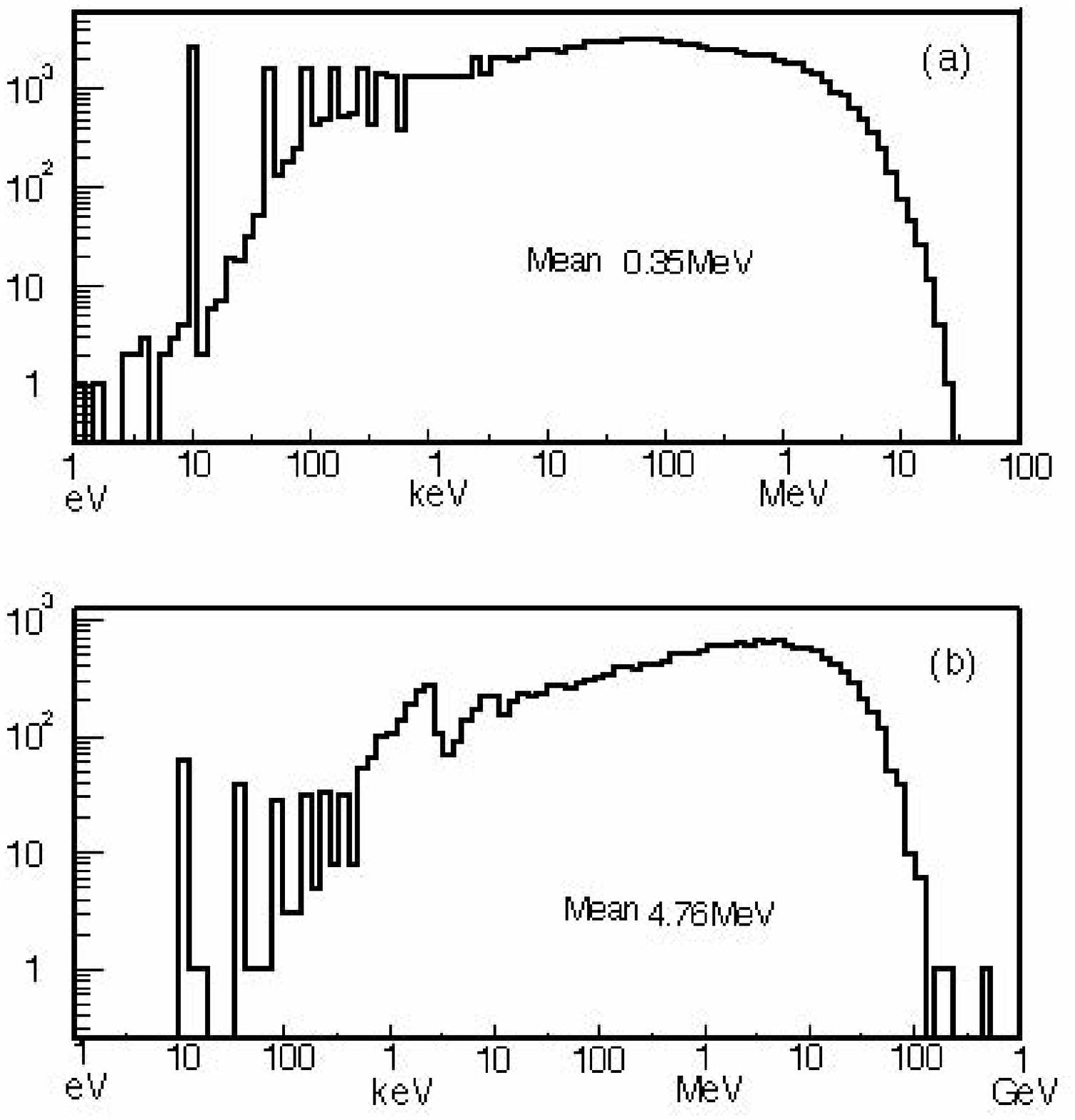}}
  \caption{\label{fig:sor_energy}The energy distributions of the synchrotron radiation
photons arising from (a) a nominal beam and (b) a beam halo. }
\end{center}
\end{figure}

Synchrotron radiation photons reached at  the IP are predominantly
generated at the final quadrupole magnets.   Photons
generated at the bending magnets upstream of the final quadrupole
magnets also reach at the IP. 
Figure~\ref{fig:sor_space}~(a) shows the
spatial distribution of the synchrotron radiation photons arising from
a nominal beam (core beam) containing $10^5$ electrons.  The extent of
$y$ distribution is limited within $\sim 0.5$~mm.  The $x$
distribution, however, is wider and has a long tail extending to $\sim
1$~cm.  This is due to photons generated at the bending magnets.
These photons are mostly contained within the beampipe (nominally with
a radius of 2~cm) even without any mask and therefore, will not be
serious background to the detector.  The energy distribution of the
photons is shown in Fig.~\ref{fig:sor_energy}~(a).  The mean photon
energy is $~\sim 0.35$~MeV.

\begin{figure}[htb]
\begin{center}
\resizebox{0.6\textwidth}{!}{\includegraphics{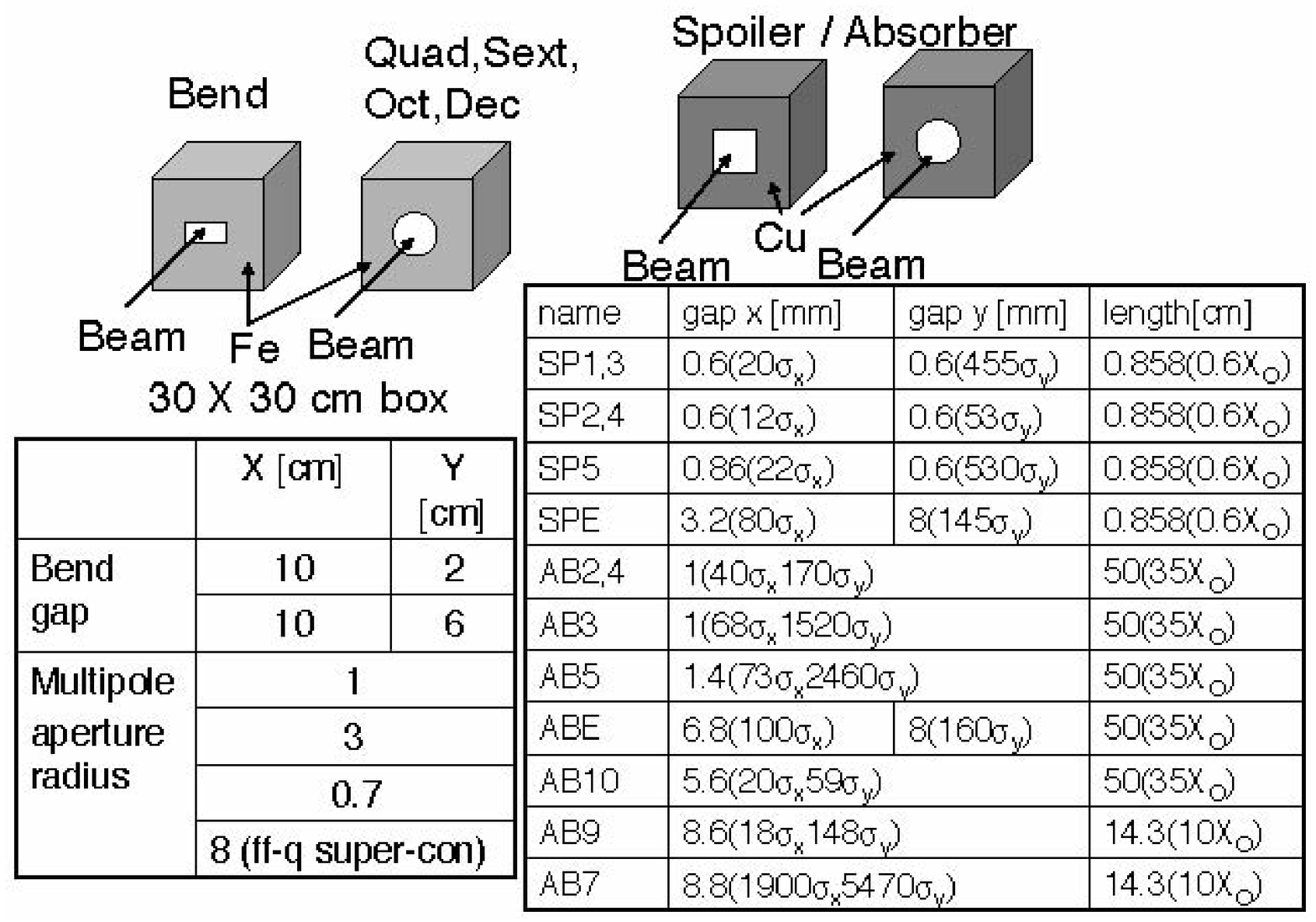}}
  \caption{\label{fig:BDScomp}Summary of dimensions and material content of BDS  
elements.}
\end{center}
\end{figure}

Synchrotron radiation photons originating from the beam halo,
electrons with large amplitudes, can be more spread and could become
serious background.
In the current design the collimator section removes halo particles
with amplitudes larger than $\sim 12\sigma_x$ or $\sim 50 \sigma_y$
from the beam center.  The size and material of spoilers and absorbers
used in our simulations are summarized in Fig.~\ref{fig:BDScomp}.
In order to simulate the collimator section performance and the
remaining synchrotron radiation, we generate the beam halo electrons
that distribute uniformly over $50 \sigma_x$ and $200 \sigma_y$ region,
where $\sigma_x$ and $\sigma_y$ are the nominal beam size at the
entrance of the BDS. The angular distribution of the halo is also
uniform over $50 \sigma_{x^\prime}$ and $200 \sigma_{y^\prime}$.  The
spatial and energy distributions of the resulting photons (without any
mask near the IP) are shown in Figs.~\ref{fig:sor_space}~(b) and
\ref{fig:sor_energy}~(b).  These photons are still contained 
in  the beampipe and  will further be 
reduced by a mask system.

\subsection{Muon Background}
\begin{figure}[htb]
    \begin{center}
      \resizebox{0.5\textwidth}{!}{\includegraphics{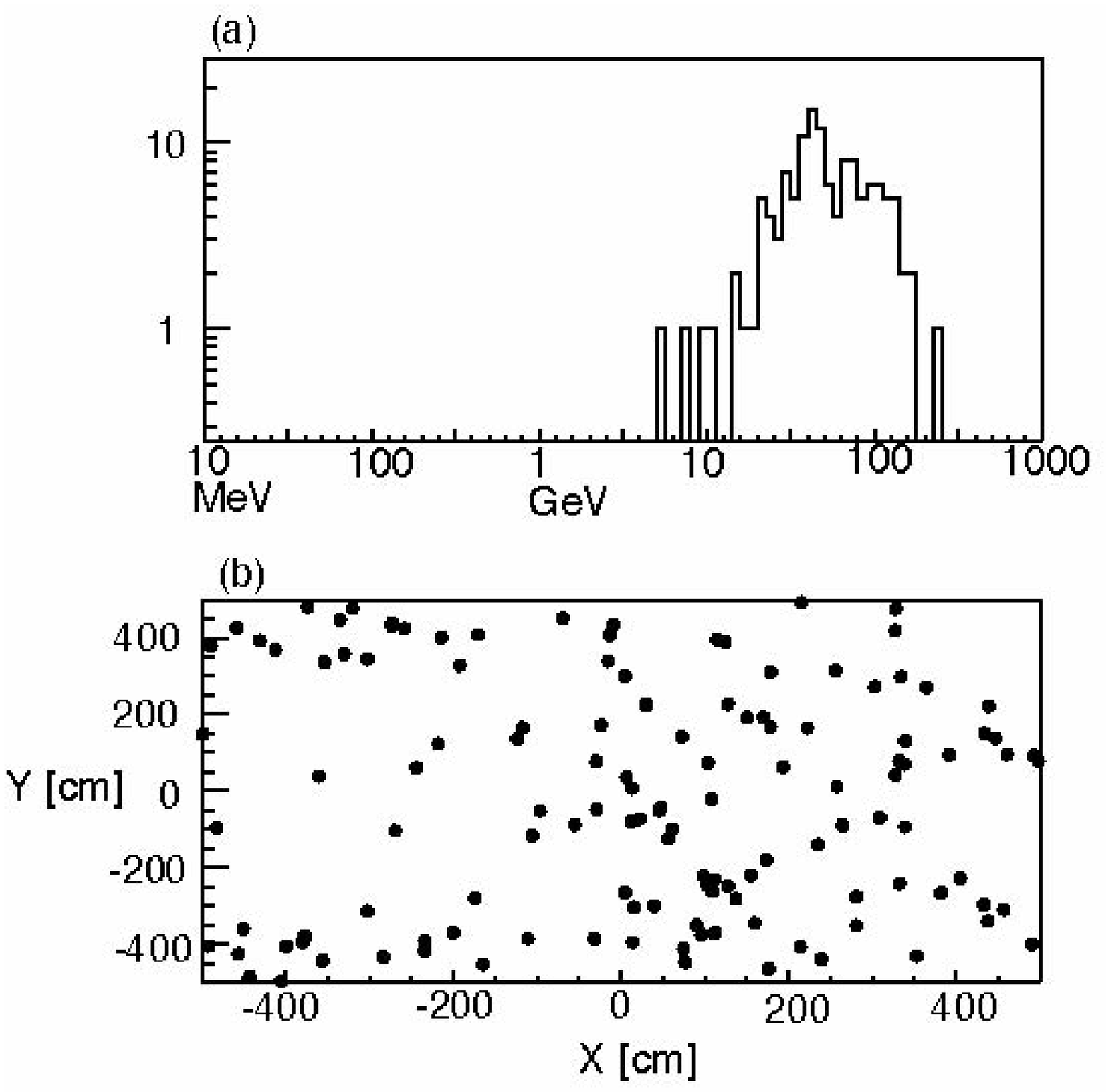}}
      \caption{The energy (a) and spatial (b) distributions of the
      muons reached at the IP if the BDS is not equipped with any muon
      shield.
      \label{fig:MuonIP}}
    \end{center}
\end{figure}

\begin{figure}[htb]
    \begin{center}
      \resizebox{0.5\textwidth}{!}{\includegraphics{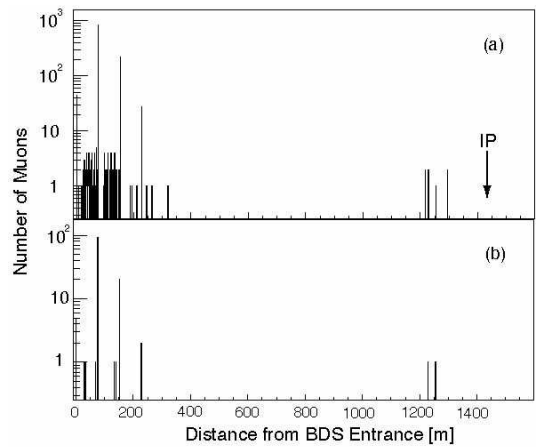}}
      \caption{Production locations for (a) all muons and (b) those
     muons which reached the IP.  \label{fig:MuonVertex}}
  \end{center}
\end{figure}

The collimator section in turn can become a source of the background.
Muon pairs can be produced via photon conversions in the
electromagnetic showers created by the lost halo electrons.
Figure~\ref{fig:MuonIP} shows the energy and spatial distributions of
the muons reached at the IP if the BDS is not equipped with any muon
shield.  The average muon energy is $~61$~GeV. The probability for a 250~GeV
electron lost at the collimator to produce a muon at the
IP (without muon shield) is estimated to be $~\sim 1.4\times 10^{-5}$.
Figure~\ref{fig:MuonVertex} shows where along the BDS those muons are
produced.  They are produced predominantly at the collimator section.
The magnetized iron pipes surrounding the beam pipe have been considered 
as an option of the muon shield.
Its design and detailed simulation are under study.

\section{Conclusion}
We have presented some of the early results obtained using a beam delivery 
system simulation based on GEANT4.
This program, LCBDS, has successfully reproduced the beam profile
as designed.
We have shown the results on the synchrotron radiation and muons produced 
in the JLC beam delivery system.


\begin{thebibliography}{99}
\bibitem {LCBDS}  Linear Collider Beam Delivery System (LCBDS) simulation code, 
M.~Iwasaki (masako@phys.s.u-tokyo.ac.jp) and K.~Tanabe (tanabe@hep.phys.s.u-tokyo.ac. jp),  
unpublished. 
\bibitem {GEANT4} http://wwwinfo.cern.ch/asd/geant4/description.html
\bibitem {SAD} SAD (Strategic Accelerator Design), K.~Oide {\it et  al.},
http://www-acc-theory. kek. jp/Accelerator/index.html 
\end{thebibliography}
\end{document}